\documentclass[prb,twocolumn,showpacs]{revtex4}        
\usepackage{epsfig,amsmath}

\begin{document}
\title{Supercurrent transferring through $c$-axis cuprate Josephson junctions with thick normal-metal-bridge}
\author{Xin-Zhong Yan,$^{1}$ and C. S. Ting$^{2}$}
\affiliation{$^{1}$Institute of Physics, Chinese Academy of Sciences, P.O. Box 603, Beijing 100190, China\\
$^{2}$Texas Center for Superconductivity, University of Houston, Houston, Texas 77204, USA}
\date{\today}

\begin{abstract}
With simple but exactly solvable model, we investigate the supercurrent transferring through the $c$-axis cuprate superconductor-normal metal-superconductor junctions with the clean normal metal much thicker than its coherence length. It is shown that the supercurrent as a function of thickness of the normal metal decreases much slower than the exponential decaying expected by the proximity effect. The present result may account for the giant proximity effect observed in the $c$-axis cuprate SNS junctions.
\end{abstract}

\maketitle


In a superconductor-normal metal junction, it is considered that the Cooper pair can penetrate into the normal metal within a distance of coherence length $\xi_n$ due to the proximity effect.\cite{degennes} Therefore, according to the proximity theory, the supercurrent cannot transfer through a superconductor-normal metal-superconductor (SNS) junction when the length of the normal metal is much longer than $\xi_n$. However, the supercurrent in high-temperature-superconductor (HTSC) junctions with very thick barrier (consisting of weakly doped nonsuperconducting cuprates) has been observed by a number of experiments.\cite{Kasai,Kabasawa,Tarutani,Kasai1,Barner,Delin,Decca,Bozovic} There have been some theoretical explanations based on the assumption of the existence of superconducting puddles in the pseudogap states of the cuprates.\cite{Kresin,Alvarez} But the physics of pseudogap states of the cuprates is not clearly understood so far. The problem whether the supercurrent can transfer through a long bridge of normal metal between two superconductors is still an outstanding puzzle. 

The supercurrent stems from the motion of paired carriers in the superconductor. It is known that the supercurrent can be conducted by Andreev reflections in the SNS junctions.\cite{Andreev,Kulik,Ishii,Bardeen,Svidzinsky,Buttiker,Furusaki,Beenakker,Furusaki1,Schussel,Hurd,Golubov} The supercurrent in one superconductor, for example in the left one, can transfer through the SN interface by generating the propagations of electrons and holes in the normal metal due to the Andreev refection. At another NS interface, the electrons and holes are converted into paired electrons in the right superconductor.\cite{BTK} As a result, the paired particles are conducted from the left to right superconductors even though Cooper pairs cannot survive in the normal metal. In case of clean normal metal without large damping in particle propagations, the supercurrent may transfer through the long SNS junction. Most of the existing calculations were based on various approximations including the Andreev and WKB ones neglecting the normal reflections.\cite{Kulik,Ishii,Bardeen,Svidzinsky} However, the Andreev approximation is not adequate for the bound states of energy within the superconducting gap.\cite{Hurd} For long normal metal junctions, the contribution from the bound states to the supercurrent is significant because the number of the bound states is proportional to the length of the normal metal. Moreover, for long normal metal SNS junctions, one needs to carefully deal with the wavefunctions of the quasiparticles since the supercurrent sensitively depends on them. A reliable study on the supercurrent in the long SNS junctions is is still lacking. 

In this work, on the basis of the Andreev-reflection approach, we study the supercurrent in the $c$-axis cuprate SNS junctions using simple but exactly solvable model. We will show that the supercurrent can transfer through the junctions with the normal metal much thicker than its coherence length. We intend to provide a possible explanation to the relevant experiments.


We consider a $c$-axis cuprate SNS junction with the normal metal occupying the layers from $1-l$ to $l-1$. A sketch of the junction is shown in Fig. 1. Within the $ab$-plane, the quasiparticles are described by the $t-t'$ tight-binding model with $t'/t = -0.3$. The phase difference $\phi$ between the pair potentials of the two superconductors drive the supercurrent. With an unitary transformation $\hat U = \exp(i\sigma_3\phi/4)$, one can show that the physical quantities of the system depend only on the total phase difference of the pair potentials. For convenience, we here set the phases of pair potentials as $\pm\phi/2$ respectively for the left and right superconductors. The electron motion along $c$-axis is described by the interlayer hopping $t_c$ with $t_c/t << 1$. The magnitude of $t_c$ may be the same order as $\Delta$. The parameter of electron hopping through the SN interface is $t_0$. Through out this paper, we use the units of $e = \hbar$ = 1 with $-e$ as the charge of an electron. 

\begin{figure} 
\centerline{\epsfig{file=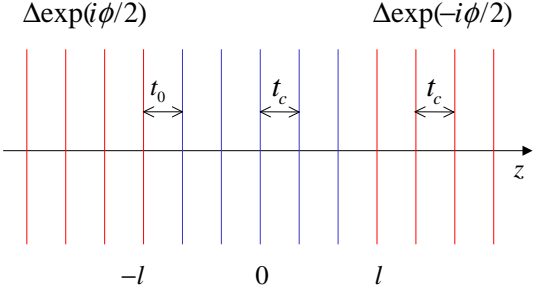,width=7.5 cm}}
\caption{(Color online) Sketch of a $c$-axis cuprate SNS junction. The interlayer hopping is described by $t_c$. $t_0$ is the hopping through the NS interface.}
\end{figure} 

The states of the quasiparticles are described by the Bogoliubov-de Gennes (BdG) equation.\cite{degennes1} Since the momentum parallel to the interfaces is conserved during the motion of the quasiparticles through the junction, the transverse (orthogonal to the $c$ axis) part of the wave function can be taken as plane waves. The problem is then reduced to solve the one-dimensional BdG equation along the $z$-direction. For an eigenstate of transverse momentum $k_{\bot}$, the chemical potential in the BdG equation is then substituted with $\tilde\mu(k_{\bot}) = \mu-\epsilon(k_{\bot})$ where $\epsilon(k_{\bot})=-2t(\cos k_1+\cos k_2)-4t'\cos k_1\cos k_2$ is the in-plane single-particle energy with $k_1$ and $k_2$ the two components of $k_{\bot}$. The order parameters of the superconductors are then given by $\Delta(k_{\bot})\exp(\pm i\phi/2)$ (+ and - for left and right superconductors, respectively) with $\Delta(k_{\bot}) = \Delta(\cos k_1 - \cos k_2)$. The BdG equation reads
\begin{equation}
\sum_jH_{ij}\psi(j) = E\psi(i), \label{bdgl}
\end{equation}
with
\begin{eqnarray}
H_{ij} = \begin{cases}[v_i-\tilde\mu(k_{\bot})]\sigma_3+\Delta_i\sigma^++\Delta^{\ast}_i\sigma^-,~~{\rm for~ } i=j\\
-t_c\sigma_3, ~~{\rm for~nearest-layer~hoppings} \\
-t_0\sigma_3, ~~{\rm for~ interface~hoppings}
\end{cases}\nonumber
\end{eqnarray}
where $v_i = V_0$ for $1-l < i < l-1$ or 0 otherwise, $\Delta_i = \Delta(k_{\bot})\exp(i\phi/2)$ for $i \leq -l$, $\Delta_i = \Delta(k_{\bot})\exp(-i\phi/2)$ for $i \geq l$, and $\sigma$'s are the Pauli matrices. The potential shift $V_0$ controls the density difference between the normal metal and the superconductors. All the states in a complete basis can be divided into three types: incoming waves of free states from the left and right superconductors, and the bound states mainly confined in the normal metal with damping tails in the two superconductors. 

The free state with an incoming wave number $k^+$ from the left superconductor is obtained as \cite{Furusaki}
\begin{eqnarray}
\psi_{l1}(j) &=& \begin{pmatrix} ue_{\phi}\\ve^{\ast}_{\phi} \end{pmatrix}(e^{ik^+z'}+ be^{-ik^+z'})+ a\begin{pmatrix} ve_{\phi}\\ue^{\ast}_{\phi} \end{pmatrix}e^{ik^-z'},
 \nonumber\\
 && ~~~~~~~~~~~~~~~~~~~~~~~~~~~~~~~~~~~ z' = j+l \leq 0 \nonumber\\
\psi_{l2}(j) &=& \begin{pmatrix} A_1e^{iq_1j}+ A_2e^{-iq_1j}\\B_1e^{-iq_2j}+ B_2e^{iq_2j} \end{pmatrix},
~~1-l \leq j \leq l-1 \nonumber\\
 \nonumber\\
\psi_{l3}(j) &=& c\begin{pmatrix} ve^{\ast}_{\phi}\\ue_{\phi} \end{pmatrix}e^{-ik^-z'} + d\begin{pmatrix} ue^{\ast}_{\phi}\\ve_{\phi} \end{pmatrix}e^{ik^+z'},
 \nonumber\\
  && ~~~~~~~~~~~~~~~~~~~~~~~~~~~~~~~~~~z' = j - l \geq 0 \nonumber 
\end{eqnarray}
with the boundary conditions
\begin{eqnarray}
r\psi_{l1}(-l) - \psi_{l2}(-l) = 0 \nonumber\\
\psi_{l1}(1-l) - r\psi_{l2}(1-l) = 0 \nonumber \\
r\psi_{l2}(l-1) - \psi_{l3}(l-1) = 0 \nonumber\\
\psi_{l2}(l) - r\psi_{l3}(l) = 0 \label{bc1}
\end{eqnarray}
where $e_{\phi}=\exp(i\phi/4)$, the wave numbers $q_1$, $q_2$,  $k^+$ and $k^-$ satisfy the equations $\xi(q_1) + V_0$ = $-\xi(q_2) - V_0$ = $\sqrt{\xi^2(k^+)+\Delta^2(k_{\bot})} \equiv E_k$ and $\xi(k^-) = -\xi(k^+)$ with $\xi(k) = -2t_c\cos k - \tilde \mu(k_{\bot})$, $u =\sqrt{1/2 + \xi(k^+)/2E_k}$, $v =\sqrt{1/2 - \xi(k^+)/2E_k}$, and $r = t_0/t_c$. The incoming wave number $k^+$ is defined in the ranges $-k_0 < k^+ < 0$ and $k_0 < k^+ < \pi$, where the group velocity $\partial E_k/\partial k$ is positive, with $k_0 = \arccos[-\tilde\mu(k_{\bot})/2t_c]$ as the `Fermi wave number' along $z$ direction. A sketch for definition of $k^+$ and $k^-$ is shown in Fig. 2. The eight coefficients $a$, $b$, $\cdots$ are determined by the boundary conditions (\ref{bc1}). Denoting
\begin{equation}
X^t = (a,b,A_1,B_1,c,d,A_2,B_2), \nonumber
\end{equation}
with superscript $t$ implying the transpose of vector $X$, we have from Eqs. (\ref{bc1}), 
\begin{equation}
MX = Z, \label{cx}
\end{equation}
where $M$ is a $8\times 8$ matrix, and $Z$ is a column vector of 8 components.
Expressing $M$ in terms of 16-block $2\times 2$ matrices, we get
\begin{eqnarray}
M = \begin{pmatrix} 
D_1(\phi) & O(-l)& 0 & O(l) \\
D_2(\phi) & rO(1-l)& 0 & rO(l-1) \\
0 & rO(l-1)& D_2(-\phi) & rO(1-l) \\
0& O(l)& D_1(-\phi)&O(-l) 
\end{pmatrix},\label{ml}
\end{eqnarray}
\begin{eqnarray}
D_1(\phi) = \begin{pmatrix} 
rve_{\phi} & rue_{\phi}\\
rue^{\ast}_{\phi} & rve^{\ast}_{\phi}
\end{pmatrix},\nonumber
\end{eqnarray}
\begin{eqnarray}
D_2(\phi) = \begin{pmatrix} 
ve_{\phi}e_- & ue_{\phi}e^{-1}_+\\
ue^{\ast}_{\phi}e_- & ve^{\ast}_{\phi}e^{-1}_+
\end{pmatrix},\nonumber
\end{eqnarray}
\begin{eqnarray}
O(l) = \begin{pmatrix} 
 -e^{l}_1 & 0  \\
 0 &   -e^{-l}_2 
\end{pmatrix},\nonumber
\end{eqnarray}
with $e_{\pm} =\exp(ik^{\pm})$. The vector $Z$ is given by
\begin{equation}
Z^t = (-rue_{\phi},-rve^{\ast}_{\phi},-ue_{\phi}e_+,-ve^{\ast}_{\phi}e_+,0,0,0,0). \nonumber
\end{equation}

\begin{figure} 
\centerline{\epsfig{file=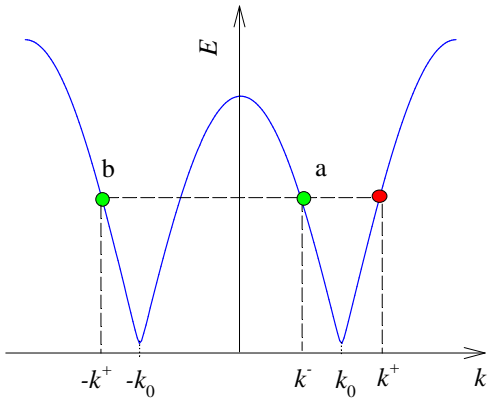,width=7. cm}}
\caption{(Color online) Sketch for definition of wave numbers $k^+$ and $k^-$ on energy curve $E(k)$. The reflected waves of an incoming wave of $k^+$ include two components of $a$ the Andreev and $b$ the normal reflections.}
\end{figure} 

Similarly, we can write down the expression for the incoming wave functions $\psi_r$ from the right superconductor. But with the configuration of the SNS junction under consideration, $\psi_r$ can be obtained from the relation
\begin{equation}
\psi_r(j;\phi) = \lambda\psi_l(-j;-\phi), 
\end{equation}
with $\lambda = \pm 1$.

For the wave function $\psi_n$ of a bound state with energy $0 < E_n < |\Delta(k_{\bot})|$, the expression is given by\cite{Hurd}
\begin{eqnarray}
\psi_{n1}(j) &=& a_n\begin{pmatrix} u^{\ast}_ne_{\phi}\\u_ne^{\ast}_{\phi} \end{pmatrix}e^{ik^{\ast}z'} + b_n\begin{pmatrix} u_ne_{\phi}\\u^{\ast}_ne^{\ast}_{\phi} \end{pmatrix}e^{-ikz'},
 \nonumber\\
 && ~~~~~~~~~~~~~~~~~~~~~~~~~~~~~~~~~~~z' = j+l < 0 \nonumber\\
\psi_{n2}(j) &=& \begin{pmatrix} A^n_1e^{iq^n_1j}+ A^n_2e^{-iq^n_1j}\\B^n_1e^{-iq^n_2j}+ B^n_2e^{iq^n_2j} \end{pmatrix},
~~-l < j < l \nonumber\\
 \nonumber\\
\psi_{n3}(j) &=& c_n\begin{pmatrix} u^{\ast}_ne^{\ast}_{\phi}\\u_ne_{\phi} \end{pmatrix}e^{-ik^{\ast}z' + }d_n\begin{pmatrix} u_ne^{\ast}_{\phi}\\u^{\ast}_ne_{\phi} \end{pmatrix}e^{ikz'},
 \nonumber\\
 && ~~~~~~~~~~~~~~~~~~~~~~~~~~~~~~~~~~~z' = j - l > 0 \nonumber 
\end{eqnarray}
where $k$ is a complex wave number determined by $\xi(k)  = i\gamma$ with $\gamma = \sqrt{\Delta^2(k_{\bot})-E_n^2}$ (Im$k > 0$), $u_n =\exp(i\theta/2)/\sqrt{2}$ with $\theta =\arctan(\gamma/E_n)$, $q^n_1$ and $q^n_2$ are determined by $\xi(q^n_1) + V_0$ = $-\xi(q^n_2) - V_0 = E_n$. The vector of the coefficients 
\begin{equation}
X^t_n = (a_n,b_n,A^n_1,B^n_1,c_n,d_n,A^n_2,B^n_2), \nonumber
\end{equation}
now satisfies the following equation
\begin{equation}
M_nX_n = 0, \label{cxn}
\end{equation}
where $M_n$ is a counterpart of $M$ with $u$, $v$, $k^+$, $k^-$, $q_1$ and $q_2$ replaced with $u_n$, $u_n^{\ast}$, $k$, $k^{\ast}$, $q^n_1$ and $q^n_2$, respectively. The energy $E_n$ is then determined by 
\begin{equation}
{\rm det} (M_n) = 0. \label{en}
\end{equation} 
The solution to the $j$th component of $X_n$ is given by the algebraic complement minor of $1j$th element of $M_n$ (multiplied by a factor that is determined by the normalization condition $\langle\psi_n|\psi_n\rangle = 1$). We note at this moment that $(a_n,b_n,A^n_1,B^n_1)$ = $\pm(c_n,d_n,A^n_2,B^n_2)$ at $\phi = 0$ because of $\psi_n(x;\phi) = \pm\psi_n(-x;-\phi)$. Therefore, for finite phase difference, the coefficients $a_n,b_n,A^n_1,B^n_1$ should have respectively the same orders of magnitudes of $c_n,d_n,A^n_2,B^n_2$. There are various approximations based on the Andreev and WKB approximation in the existing theories.\cite{Kulik,Ishii,Bardeen,Svidzinsky} The approximation in Ref. \onlinecite{Bardeen} corresponds to $b_n = c_n = A^n_2 = B^n_1 = 0$, taking into account only the Andreev reflections but neglecting the normal reflections at the right NS interface. It is not correct. Actually, in a bound state, the electrons and holes are bounced back and forth again and again in the normal metal. The normal and Andreev reflections at the two interfaces are equally important. 

To derive the expression of the current, we start from the operator of current density in the continuum model of normal metal
\begin{equation}
J(x) = -{\rm Im}[\psi^{\dagger}(x)\nabla\psi(x)]. \label{jo}
\end{equation}
For the lattice case, $\nabla\psi(x)$ in Eq. (\ref{jo}) is replaced with $[\psi(j+1)-\psi(j-1)]/2$.
Taking the statistical average with summing up all the contributions from the states of positive and negative energies, we obtain
\begin{eqnarray}
J&=& \int_{BZ}\frac{d\vec k_{\bot}}{(2\pi)^2}[\int\frac{dk^+}{2\pi}\tanh(\frac{E_k}{2T})J_f(\vec k)
 \nonumber\\
 & & +\sum_n \tanh(\frac{E_n}{2T}){\rm Re}(A^{n\ast}_+A^{n}_-\sin q^n_1-B^{n\ast}_+B^{n}_-\sin q^n_2)]\nonumber\\
  \label{J}
\end{eqnarray}
with
\begin{eqnarray}
J_f(k)={\rm Re}\{[A^{\ast}_+(\phi)A_-(\phi)-A^{\ast}_+(-\phi)A_-(-\phi)]\sin q_1  \nonumber\\
 -[B^{\ast}_+(\phi)B_-(\phi)-B^{\ast}_+(-\phi)B_-(-\phi)]\sin q_2\}. \nonumber
\end{eqnarray}
where the integral $\int_{BZ}d\vec k_{\bot}$ runs over the first Brillouin zone, $A_{\pm} = A_1 \pm A_2$, and $T$ is the temperature. The first term $J_f$ in right hand side of Eq. (\ref{J}) comes from the contributions of the free states. The second term is due to the bound states. Here, the phase dependence of the coefficients of the free waves is explicitly indicated by $\phi$ as their argument. Of course, those coefficients of the bound states $A^n$'s and $B^n$'s, the energy $E_n$ and the wave numbers $q^n$'s depend on the phase $\phi$ as well. At $\phi =0$ corresponding to the equilibrium state, there is no current flowing through the junction. The current is driven by a finite phase difference. Instead to investigate the phase dependence, we here confine ourselves to the problem of length $L = 2l$ dependence of the supercurrent with fixed phase difference $\phi = \pi/2$.

We have used two sets of the parameters $(t_c,~\Delta,~T)/t$ = $(5,~4.75,~0.25)\times 10^{-2}$ (S1) and $(6,~3.8,~0.2)\times 10^{-2}$ (S2) in our calculation. The parameter of electron hopping through the SN interface was chosen as $t_0 = 0.8 t_c$. The chemical potential $\mu$ and the potential shift $V_0$ in the normal metal were set respectively to $\mu/t = -0.97$ and $V_0/t = -0.042$, corresponding to the hole densities $\delta_s \approx 0.13$ in the superconductor and $\delta_n \approx 0.11$ in the normal metal (at finite temperature). Shown in Fig. 3 are the calculated results for the supercurrent as function of the distance  $L=2l$ (in unit of $c$-direction lattice constant) between two superconductors. The circles and squares correspond to the parameters of (S1) and (S2), respectively. For comparison, we also depict the curves of $J(L) \propto \exp(-L/\xi)$ with $\xi$ = 4 and $\xi$ = 6. By the proximity theory, $J(L)$ should decay exponentially with a much shorter coherence length $\xi_n \sim 1 - 2$ \AA~ for cuprates. Our result implies that the supercurrent can flow through the junction with $L > \xi >> \xi_n$.

\begin{figure} 
\centerline{\epsfig{file=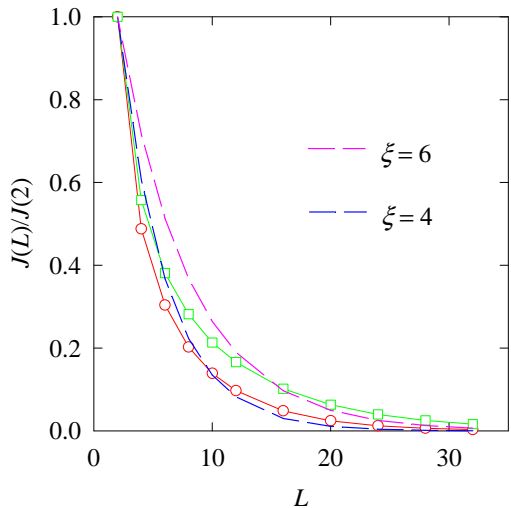,width=7.5 cm}}
\caption{(Color online) $c$-axis supercurrent $J$ through a $d$-wave SNS junction as function of the distance $L$ between two superconductors. $L$ is given in unit of $c$-axis lattice constant. The circles and squares are the calculated results for two sets of parameters (see the text). The dashed lines express the formula $\exp[-(L-2)/\xi]$ with $\xi$ = 4 and $\xi$ = 6, respectively.}
\end{figure} 

To compare the present calculation with the proximity theory for the layered system, we here estimate the theoretical coherence length $\xi_c$ along $c$ axis \cite{Delin}. According to the uncertainty principle, $\xi_c$ is proportional to the inverse of the uncertainty of momentum $\delta p_c$ of electrons. The latter can be estimated as $v_c\delta p_c \approx \delta E$, where $v_c$ is the averaged magnitude of the electron velocity along $c$-direction. Note that there is no Fermi surface across the $c$-axis in the layered system with weak interlayer hopping. From the energy dispersion in $c$-direction, $\epsilon(q) = -2t_c\cos(q)$ with $q$ the $c$-axis momentum in unit of $c$ ($c$-axis lattice constant) = 1, we obtain the electron velocity $2t_c\sin(q)$. The overall magnitude of $v_c$ can be estimated as $t_c$. On the other hand, the uncertainty of the energy $\Delta E$ is the order of the bandwidth $4t_c$. We then have $\xi_c \approx 1/4$. This $\xi_c$ is approximately the same as the observed data for cuprates. Therefore, according to the proximity theory, the supercurrent cannot transfer along the $c$ axis even for very short SNS junctions. However, since the supercurrent can be conducted by the Andreev reflections, it is not limited by the coherence length. Our calculation may account for the giant proximity effect observed in the cuprates SNS junctions. 


In summary, we have investigated the supercurrent transferring through the $c$-axis cuprate SNS junctions. Due to the Andreev reflections, the supercurrent is conducted by the in-gap bound states and the free states above but close to the gap. It is shown that the supercurrent as a function of thickness of the normal metal decreases much slower than the exponential decaying expected by the proximity effect. This result implies that the supercurrent can transfer through the clean $c$-axis cuprate SNS junctions with the normal metal much thicker than its coherence length. The present result may account for the giant proximity effect observed in the cuprate SNS junctions. 


This work was supported by the National Basic Research 973 Program of China under grant No. 2005CB623602, and NSFC under grant No. 10774171, a grant from the Robert A. Welch Foundation under No. E-1146, and the TCSUH.

\end{document}